\documentclass[aps,prx,reprint,superscriptaddress]{revtex4-1}
\usepackage{lmodern}
\usepackage{graphicx}
\usepackage{braket}
\usepackage{amsfonts, amsmath, amsthm, amssymb}
\usepackage{bbold}
\usepackage{subfigure}
\usepackage{hyperref}

\hypersetup{colorlinks=true,linkcolor=blue,citecolor=blue,urlcolor=blue}
\def\equationautorefname~#1\null{(#1)\null}
\usepackage{cleveref}
\usepackage{ulem}
\usepackage[utf8]{inputenc}
\usepackage[T1]{fontenc}
\Crefname{figure}{Fig.}{Figs.}
\newcommand*{\eqnref}[1]{%
\begingroup
Eq. \autoref{#1}%
\endgroup}
\makeatletter
\newsavebox{\@brx}
\newcommand{\llangle}[1][]{\savebox{\@brx}{\(\m@th{#1\langle}\)}%
  \mathopen{\copy\@brx\kern-0.5\wd\@brx\usebox{\@brx}}}
\newcommand{\rrangle}[1][]{\savebox{\@brx}{\(\m@th{#1\rangle}\)}%
  \mathclose{\copy\@brx\kern-0.5\wd\@brx\usebox{\@brx}}}
\makeatother

\begin{document}

\title{Floquet vortex states induced by light carrying the orbital angular momentum}

\author{Hwanmun Kim}
\affiliation{Joint Quantum Institute, NIST and University of Maryland, College Park, Maryland 20742, USA}
\affiliation{Department of Physics, University of Maryland, College Park, Maryland 20742, USA}

\author{Hossein Dehghani}
\affiliation{Joint Quantum Institute, NIST and University of Maryland, College Park, Maryland 20742, USA}
\affiliation{Departments of Electrical and Computer Engineering and Institute for Research in Electronics and Applied Physics, University of Maryland, College Park, Maryland 20742, USA}

\author{Iman Ahmadabadi}
\affiliation{Joint Quantum Institute, NIST and University of Maryland, College Park, Maryland 20742, USA}
\affiliation{Department of Physics, University of Maryland, College Park, Maryland 20742, USA}

\author{Ivar Martin}
\affiliation{Materials Science Division, Argonne National Laboratory, Argonne, Illinois 60439, USA}

\author{Mohammad Hafezi}
\affiliation{Joint Quantum Institute, NIST and University of Maryland, College Park, Maryland 20742, USA}
\affiliation{Department of Physics, University of Maryland, College Park, Maryland 20742, USA}
\affiliation{Departments of Electrical and Computer Engineering and Institute for Research in Electronics and Applied Physics, University of Maryland, College Park, Maryland 20742, USA}

\date{\today}

\begin{abstract}
We propose a scheme to create an electronic Floquet vortex state by irradiating a two-dimensional semiconductor with the laser light carrying non-zero orbital angular momentum. We analytically and numerically study the properties of the Floquet vortex states, with the methods analogous to the ones previously applied to the analysis of superconducting vortex states. We show that such Floquet vortex states are similar to the superconducting vortex states, and they exhibit a wide range of tunability. To illustrate the potential utility of such tunability, we show how such states could be used for quantum state engineering.
\end{abstract}

\pacs{}
\maketitle


\textit{Introduction.--} Quantum vortices and localized quantum states associated with them have long a subject of active interest in diverse areas of physics \cite{onsager1949statistical,FEYNMAN195517,abrikosov1957magnetic, CAROLI1964307, JACKIW1981681, volovik2003universe}. To create and observe such quantum vortex states, numerous efforts have been made in diverse systems such as Bose-Einstein condensates \cite{donnelly1991quantized,PhysRevLett.84.806,Abo-Shaeer476}, superconductors \cite{RevModPhys.66.1125,RevModPhys.73.867}, and magnetic materials \cite{BOGDANOV1994255,roessler2006spontaneous,nagaosa2013topological}. While the quantum vortex states themselves exhibit many exotic quantum and classical many-body phenomena \cite{Kosterlitz_1973, Kosterlitz_1974, PhysRevLett.115.010401, PhysRevB.43.130, PhysRevLett.61.1666, PhysRevLett.63.1511}, their stability as topological defects makes them a promising quantum platform for applications such as quantum information processing \cite{RevModPhys.80.1083,PhysRevB.73.014505,PhysRevLett.95.173601}.

Recently, Floquet systems have become popular as a useful way to engineer exotic quantum states \cite{PhysRevB.79.081406,lindner2011floquet,rechtsman2013photonic,PhysRevLett.110.016802,PhysRevB.88.224106,PhysRevX.4.031027,PhysRevX.6.021013,PhysRevB.90.195429,PhysRevResearch.2.043004, McIver2020, Sato2019}. Moreover, there have been many recent advancements in the spatial control of optical beams in atomic systems \cite{Zupancic:16,Barredo1021,barredo2018synthetic,schine2019electromagnetic}. These techniques have the potential to be applied to  electronic systems and can provide a wide range of tunability in quantum state engineering. 

\begin{figure}[t]
\centering
\includegraphics[width=\linewidth]{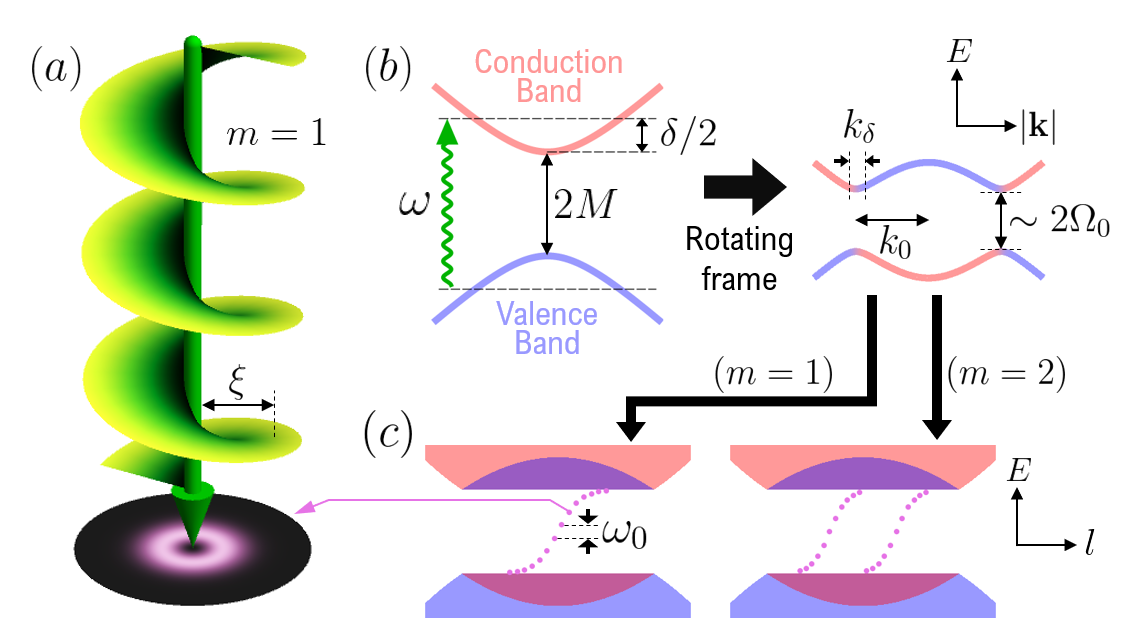}\subfigure{\label{F1a}}\subfigure{\label{F1b}}
\caption{(a) A 2D semiconductor illuminated by a laser light carrying OAM. The applied light field has the optical vortex structure of size $\xi$. The figure illustrates the case of vorticity $m=1$. (b) The laser field has frequency $\omega$, and couples the conduction and the valence bands of the semiconductor with the gap $2M$. The detuning is $\delta = \omega-2M$. In the rotating frame, the hybridization gap of about $2\Omega_0$ develops around the resonance ring whose radius and thickness are $k_0$ and $k_\delta$, respectively. (c) For the light field with non-zero vorticity $m$, $|m|$ branches of Floquet vortex states develop in the middle of the hybridization gap. Around the zero energy, each branch has linear dispersion with energy separation $\omega_0$ between nearby states in the branch. Note that the energy spectrum is illustrated with respect to the electronic pseudo-OAM, $l$.}\label{F1}
\end{figure}

In this paper, we present a scheme to create Flouqet quantum vortex states by shining a light field carrying orbital angular momentum (OAM) on a two-dimensional (2D) semiconductor, as illustrated in Fig.\ref{F1}. In small detuning and the weak field limit, we show that electronic Floquet vortex states are localized around the optical vortices with localization length bounded by the shape and intensity of the optical field. We also show that the number of vortex state branches is directly given by the vorticity of the light, which quantifies the OAM carried by each photon. Such close relation with OAM of light distinguish these vortex states from the edge states of the uniform Floquet Chern insulator \cite{lindner2011floquet} or the vortex states introduced in Ref. \cite{PhysRevLett.110.016802,PhysRevB.88.224106}. While many characteristics of these Floquet vortex states carry close analogy with superconducting systems, we show that the Floquet vortex states in the current system benefit from a very broad range of tunability. For example, the freedom to choose the size of the optical vortex can be used as a knob to control the non-linearity of the vortex state spectrum. To demonstrate how such tunability can be exploited for quantum state engineering, we construct a scheme of quantum information processing based on optically manipulating Floquet vortex states,  with simple single-qubit and two-qubit operations.


\textit{Model.--} We consider $H_0 = (vk_x, vk_y, M)\cdot\sigma$ as our model for a spinless 2D semiconductor \cite{Bernevig1757,PhysRevB.72.035321}. For brevity, we have set $\hbar=1$. Here, $\sigma=(\sigma_x,\sigma_y,\sigma_z)$ are Pauli matrices. $M$ is a half of the band gap and $v$ is a parameter determining the curvature of the band dispersion $\pm\sqrt{M^2+v^2(k_x^2+k_y^2)}$, where the positive (negative) energy states correspond to the conduction (valence) band.
We vertically shine a linearly-polarized laser field with a non-zero orbital angular momentum (OAM), $\mathbf{\mathcal{A}}(\mathbf{r},t)= A(\mathbf{r})e^{i\omega t}\mathbf{\hat{x}}+\text{c.c.}$ 
on a semiconductor, as illustrated in \Cref{F1} where $\omega$ is the frequency of the laser field. The OAM of the laser field is represented in the azimuthal phase factor of $A(\mathbf{r})=A_0(r) e^{im\phi}$, where $r=\sqrt{x^2+y^2}$ and $\phi=\arctan(y/x)$. The integer $m$ here is the vorticity of the field, and we refer the vortex structure with non-zero vorticity in the light field as an optical vortex. Due to this vortex structure, $A_0(r)$ should vanish at $r=0$. We set the size of optical vortex to $\xi$, which means that $A_0(r)$ smoothly saturates to $A_\text{max}$ at $r\ge\xi$.
With the minimal coupling $\mathbf{k}=(k_x,k_y)\to\mathbf{k} + e\mathbf{\mathcal{A}}(\mathbf{r},t)$, we obtain the time-periodic Hamiltonian
\begin{eqnarray}\label{Ht}
H(t) = H_0 + ev\mathbf{\mathcal{A}}(\mathbf{r},t)\cdot\sigma.
\end{eqnarray}

When $\omega>2M$, the frequency detuning $\delta = \omega - 2M$ becomes positive and the conduction and valence bands become resonant at the resonance ring of momentum, $|\mathbf{k}|=k_0= v^{-1}\sqrt{\omega^2/4-M^2}$.
From \eqnref{Ht}, the applied laser field generates position-dependent Rabi frequency $\Omega(r) = evA_0(r)$ and hybridizes the conduction and valence bands while opening an energy gap about $2\Omega_0$ around the resonance ring, where $\Omega_0 = \lim_{r\to\infty}\Omega(r)$. To describe these hybridized bands, we consider the transformation into the rotating frame, $U(t) = P_{c} e^{-i\omega t/2} + P_{v} e^{i\omega t/2}$, where $P_{c}$ ($P_{v}$) is the projection operator into the conduction (valence) band. In the weak field limit $\Omega_0 \ll \sqrt{\omega\delta}$,
we can drop the fast oscillating terms from the rotated Hamiltonian $-iU^\dag(t)\partial_t U(t) + U^\dag(t)H(t)U(t)$ and obtain the effective Hamiltonian under the rotating wave approximation (RWA). Furthermore, we consider the small detuning regime $\delta \ll \omega$. In this regime, we can write $\delta \simeq v^2 k_0^2 /M$ and $v k_0 \ll M$.
Then, for the small momenta $|\mathbf{k}|=O(k_0)$ \cite{supplement},
\begin{eqnarray}\label{H_rwa}
H_\text{RWA} = \frac{\delta}{2}\left(\frac{\mathbf{k}^2}{k_0^2}-1\right)\sigma_z + \left[\Omega(r) e^{-im\phi}\sigma_+ +\text{H.c.}\right],\quad
\end{eqnarray}
where $\sigma_{\pm}=(\sigma_x\pm i\sigma_y)/2$.




\textit{Floquet vortex states.--} Because of the breaking of the translational symmetry by the optically-induced vortex, it is possible to have electronics states with energies inside the spectral gap that are localized in the vicinity of the vortex.
From \eqnref{H_rwa}, we can estimate the spatial extent of such states. First, one can readily observe that the diagonal components are dominant over off-diagonal elements for most of $\mathbf{k}$s except the vicinity of the resonance ring. This means that the hybridization mostly occurs at the momenta in the narrow region near the resonance ring, and the thickness of this region can be estimated by finding the range of $|\mathbf{k}|$ that makes the off-diagonal elements of \eqnref{H_rwa} comparable to or larger than the diagonal elements. We find that the hybridization of the two bands occurs at $|\mathbf{k}|-k_0 = O(k_\delta)$ where $k_\delta \equiv k_0 \Omega_0 /\delta$, that characterizes the momentum range over which the Rabi frequency and dispersion of Eq.\ref{H_rwa} are comparable around the resonant momentum ring. If any intragap state develops within this hybridization gap, such a state should be a superposition of the Bloch states within this momentum region. Therefore $k_\delta^{-1}$ serves as a lower bound for the spatial size of such intragap state.  If a localized intragap state develops around the optical vortex, this state cannot extend to the region where $A_0(r)$ saturates to $A_\text{max}$ since the field is nearly uniform and therefore the system remains gapped. Therefore such a localized intragap state has an upper bound $O\left(k_\delta^{-1} + \xi\right)$ for its size.

\begin{figure}[t]
\centering
\includegraphics[width=1.0\linewidth]{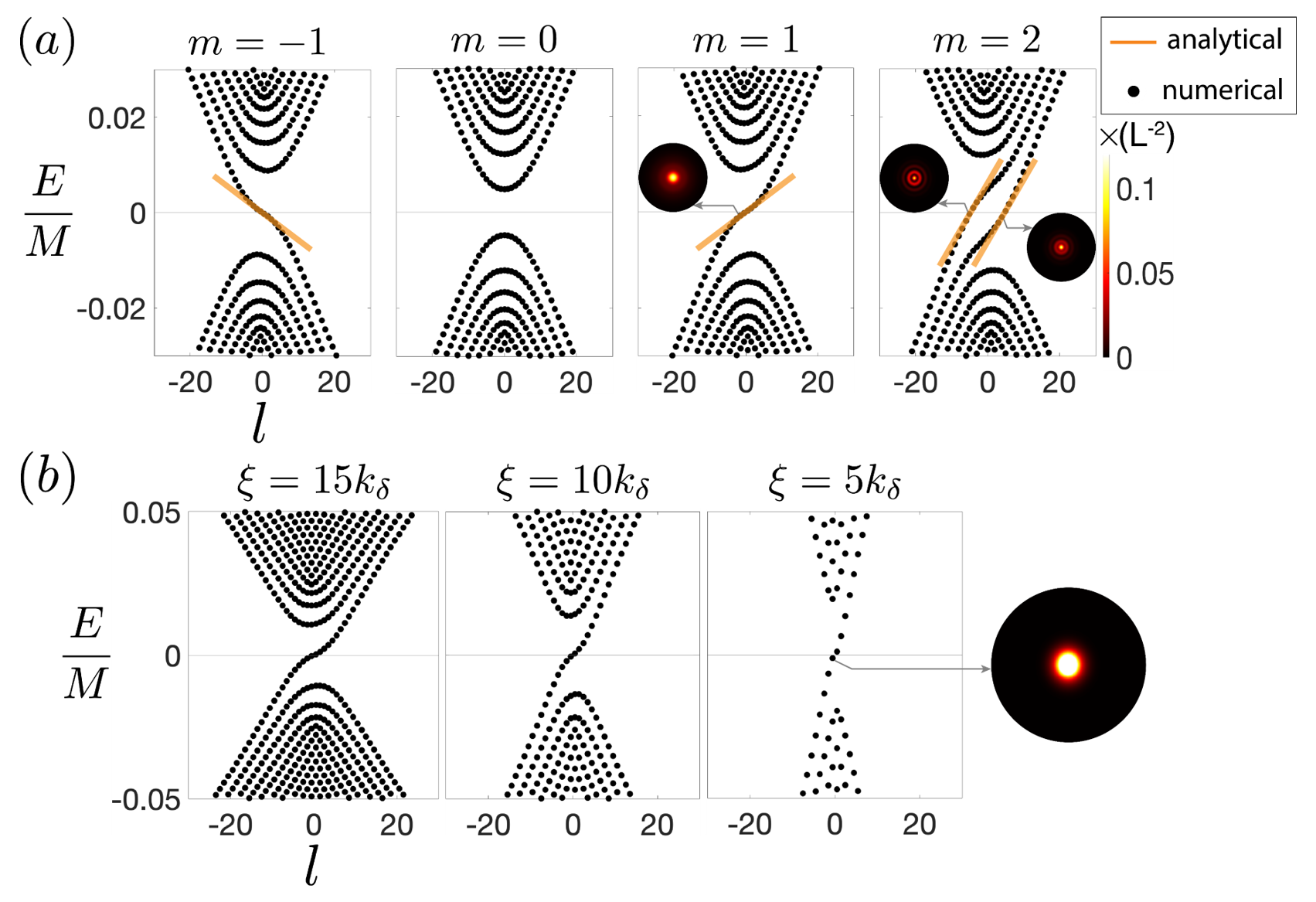}\subfigure{\label{F2a}}\subfigure{\label{F2b}}\subfigure{\label{F2c}}
\caption{(a) Numerically calculated energy spectra in terms of pseudo-OAM $l$. We use $\omega=2.05M$, $A_\text{max}=0.09M (ev)^{-1}$, and $A_0(r)=A_\text{max}\left[1-\exp\lbrace-r^2/(2\xi^2)\rbrace\right]$, $\xi=20k_{\delta}$, and suppose a disk sample of radius $25\xi$. The numerical spectra agree with the analytically expected dispersion in \eqnref{lin_disp} including the number of intragap state branches and the slope of the linear dispersion for small $|E_l|$ and $l$. Electronic density profiles of selected states are presented in the insets. (b) Dispersions for $m=1$ with identical parameters with (a) except the optical vortex size $\xi$ and the disk size $500k_{\delta}$. As $\xi$ reduces, the linear region of the spectrum shrinks while the energy separation between the nearby states increases.}\label{F2}
\end{figure}

By using the semiclassical argument introduced in Ref. \cite{volovik1993vortex}, one can show that $|m|$ branches of intragap states develop around the optical vortex with vorticity $m$ \cite{supplement}. We call these states Floquet vortex states, and we can obtain a fully quantum-mechanical description of the dispersion and wavefunction of these states by applying mathematical methods used for superconducting vortices \cite{caroli1964bound,PhysRevLett.114.195301,PhysRevB.93.174505,PhysRevLett.119.067003}. To do so, we note that while the effective Hamiltonian in \eqnref{H_rwa} does not commute with the electronic OAM, $\hat{L}=-i\partial_\phi$, it does commute with the electronic pseudo-OAM, $\hat{l}=-i\partial_\phi+(m/2)\sigma_z$. Then the eigenstates of this effective Hamiltonian can be written in the form of vortex states,
\begin{eqnarray}\label{psi_angular}
\psi_{n,l}(\mathbf{r})=\left( e^{i(l-m/2)\phi}u_{n,l,+}(r) , e^{i(l+m/2)\phi}u_{n,l,-}(r) \right)^T. \quad
\end{eqnarray}
Here, the branch index $n = 1,\cdots,m$ represents different branches of Floquet vortex states.
One can also show that this system satisfies the particle-hole symmetry which requires $\psi_{n,-l}(\mathbf{r})=i\sigma_y \psi_{|m|+1-n,l}^*(\mathbf{r})$ and $E_{n,-l}=-E_{|m|+1-n,l}$, where $E_{n,l}$ is the corresponding eigenenergy for $\psi_{n,l}(\mathbf{r})$. In the large optical
vortex regime $k_\delta^{-1}\ll\xi$,
the low-energy spectrum of these Floquet vortex states are given by \cite{PhysRevLett.119.067003}
\begin{eqnarray}\label{lin_disp}
E_{n,l} &=& ml\omega_0 +[n-(|m|+1)/2]\tilde{\omega}_0, \text{ where} \nonumber\\
\omega_0 &=& \frac{\delta \int_0^\infty \frac{\Omega(r)}{r} e^{-(2k_0/\delta)\int_0^r \Omega(r') dr'} dr}{k_0\int_0^\infty e^{-(2k_0/\delta)\int_0^r \Omega(r') dr'} dr}, \nonumber\\
\tilde{\omega}_0 &=& \frac{\delta (\pi/2)}{k_0\int_0^\infty e^{-(2k_0/\delta)\int_0^r \Omega(r') dr'} dr}.
\end{eqnarray}
Here, the energy separation between nearby states and branches, $\omega_0$ and $\tilde{\omega}_0$, respectively, are solely determined by the bulk properties and the details of the radial beam profile $A_0(r)$. These parameters are independent of the system size and therefore the energy separation between states remains in the thermodynamic limit.
This analytic expression of the dispersion is valid for the low-energy and the low-$l$ regime, $|E_{n,l}| \ll \Omega_0$ and $|l|\ll\sqrt{\delta/\Omega_0}$. \Cref{F2a} presents how this analytically found dispersion agrees with the numerical dispersion obtained by diagonalizing \eqnref{H_rwa} \cite{supplement}. As shown in the figure, the number of intragap state branches is given by $|m|$. The analytic dispersion and the numerical dispersion agree for the low-energy and low-$l$ regime, and deviate from each other as the energy or $l$ moves away from zero. Nevertheless, we can still use \eqnref{lin_disp} to get a rough estimate of the pseudo-OAM differences between different intragap state branches, in the large optical vortex regime \cite{supplement}. Assuming the entire intragap state branches are linearly dispersing, the different branches at the same energy would have the pseudo-OAM momentum difference of  $\tilde{\omega}_0/\omega_0 = O\left(k_0 k_\delta^{-1} \sqrt{k_\delta \xi}\right)$. This large difference in the angular momentum prevents the vortex modes from different branches to hybridize each other. With the same assumption, the number of states in a single branch can be also estimated as $2\Omega_0 / \omega_0 = O(k_0 \xi)$.

Note that these Floquet vortex states around the optical vortex are distinguished from the edge states of topological Floquet Chern insulators \cite{lindner2011floquet} or the vortex states introduced in Ref. \cite{PhysRevLett.110.016802,PhysRevB.88.224106}. 
For the edge state of the Floquet Chern insulator to develop, the bulk part of the system should have a non-zero Chern number, while the Floquet vortex states we are discussing appear regardless of the Chern number of the system. This point becomes clear by investigating the system under irradiation of a circularly-polarized light beam which also carries a non-zero OAM \cite{supplement}. While the bulk part of such system becomes a Floquet Chern insulator as explained in Ref. \cite{lindner2011floquet}, there are still $|m|$ branches of Floquet vortex states in the middle of the hybridization gap. The Floquet vortex states in our system also differ from the vortex states in Ref. \cite{PhysRevLett.110.016802,PhysRevB.88.224106} where the vortex structure does not couple with the electronic kinetic terms and has no trivial way to realize in experiments.

While many properties of the Floquet vortex states can be analyzed with the similar techniques used for superconducting vortex states, our Floquet vortex states have wider tunability due to the freedom to control the size of optical vortices.
For superconducting vortex states, the size of vortices is tied to $O\left(k_\delta^{-1}\right)$ since the BdG equation should be satisfied in a self-consistent way. However, \eqnref{H_rwa} does not have such constraints and we have the freedom to choose the size of the optical vortex. To illustrate the consequence of this freedom, we display the numerical dispersion for different optical vortex sizes in \Cref{F2b}. As shown in the figure, as the optical vortex size $\xi$ gets smaller, the linear region of the spectrum shrinks and therefore the non-linearity of the spectrum is enhanced. This adjustable non-linear dispersion of Floquet vortex states invites the possibility of using them as a  platform for quantum state engineering. 

\begin{figure}[t]
\centering
\includegraphics[width=\linewidth]{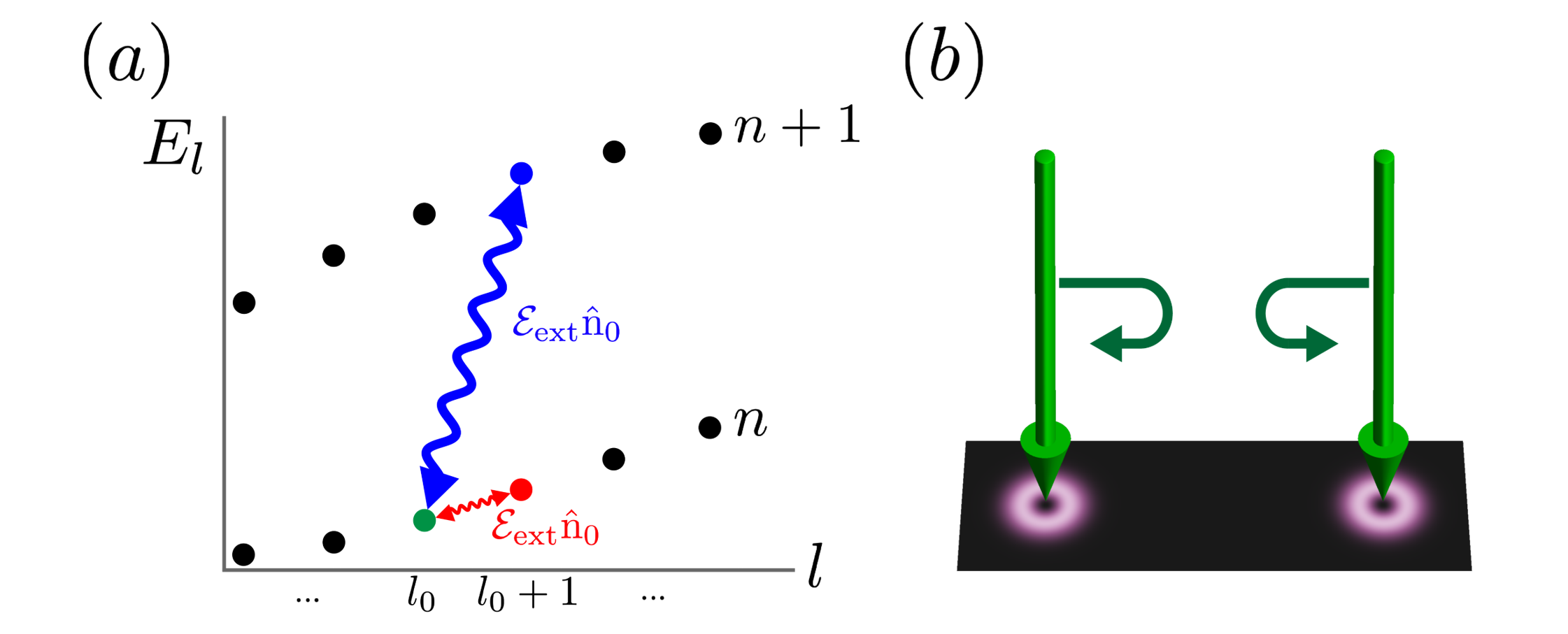}\subfigure{\label{F3a}}\subfigure{\label{F3b}}
\caption{(a) The non-linearity of the dispersion allows one to encode different Floquet vortex states as qubits. For example, the vortex states with pseudo-OAM $l_0$ and $l_0+1$ from the vortex state branch with index $n$ (red arrow) or the branches with indices $n$ and $n+1$ can be used to encode a qubit (blue arrow). Arbitrary single-qubit rotation can be performed by shining an extra linearly polarized light. While the polarization $\mathbf{\hat{n}_0}$ determines the rotation axis, the beam amplitude $\mathcal{E}_\text{ext}$ and the irradiation time determines the rotation angle. (b) Two-qubit gates can be performed by bringing two vortices close to each other and then separating them back.}\label{F3}
\end{figure}

\textit{Quantum information processing with Floquet vortex states.--} To illustrate the potential utility of the Floquet vortex states as a platform for quantum state engineering, we show how one and two-qubit operations can be performed in this system.
As we have seen in the previous section, we can increase the energy level spacing and the spectral non-linearity by reducing the size of the optical vortex. It is this enhanced non-linearity that allows to create qubits out of the Floquet vortex states and manipulate them (\Cref{F3}).

Specifically,  we consider two Floquet vortex states with pseudo-angular momentum $l_0$ and $l_0+1$ of an intragap branch with index $n$.
That is, $\braket{\mathbf{r}|0}\equiv\psi_{n,l_0}(\mathbf{r})$ and $\braket{\mathbf{r}|1}\equiv\psi_{n,l_0+1}(\mathbf{r})$. (While here we choose the vortex states from the same intragap branch, alternatively vortex states from different branches can be also used.) To manipulate this qubit, we may apply an extra linearly-polarized field to create an oscillating potential
\begin{eqnarray}
V_\text{ext}(t) = e\mathcal{E}_\text{ext} \mathbf{\hat{n}_0}\cdot\mathbf{r} \cos(\Omega_\text{ext} t),
\end{eqnarray}
where $\mathcal{E}_\text{ext}$ is the amplitude of the applied electric field and $\mathbf{\hat{n}_0}=\cos\phi_0\mathbf{\hat{x}}+\sin\phi_0\mathbf{\hat{y}}$ is the polarization of the field.
Then, in the rotating frame with frequency $\Omega_\text{ext}$, 
the effective Hamiltonian for this qubit space becomes
\begin{eqnarray}
&& H_\text{1-qubit}
= \left(E_{l_0}+\frac{\Omega_\text{ext}}{2}\right)\ket{0}\bra{0} + \left(E_{l_0+1}-\frac{\Omega_\text{ext}}{2}\right)\ket{1}\bra{1} \nonumber\\
&&\qquad\qquad + \left[ e\mathcal{E}_\text{ext} \braket{1| r\cos(\phi-\phi_0)|0}\ket{1}\bra{0} +\text{H.c.} \right], \nonumber\\
&& \braket{1| r\cos(\phi-\phi_0)|0} =\int d^2\mathbf{r} \psi_{n,l_0+1}^\dag(\mathbf{r}) r\cos(\phi-\phi_0) \psi_{n,l_0}(\mathbf{r}) \nonumber\\
&& \qquad\qquad\quad =\pi e^{i\phi_0} \sum_{s=\pm}
\int_0^\infty u^*_{n,l_0+1,s}(r) u_{n,l_0,s}(r) r^2 dr.
\end{eqnarray}
By setting $\Omega_\text{ext}=E_{n,l_0+1}-E_{n,l_0}$, we can effectively tune $H_\text{1-qubit}$ to be a superposition of $\sigma_x$ and $\sigma_y$ with an arbitrary ratio between them. Then this extra field implements an arbitrary single-qubit rotation where the rotation angle is tuned by the field amplitude $\mathcal{E}_\text{ext}$ and the irradiation time, while the rotational axis is set by the polarization $\mathbf{\hat{n}_0}$. Note that this qubit is isolated from other vortex states because the field with frequency matched to the energy difference $E_{n,l_0+1}-E_{n,l_0}$ cannot couple to other modes due to the non-linear dispersion of the vortex states. \cite{supplement} 

For two-qubit operations, we can move two vortices close to one another. This will lead to a hybridization, $J$, between the modes with the same quantum numbers on the two vortices. Yet, single-electron hopping from one vortex to another may be energetically unfavorable due to the on-site interaction energy $U$. This will generate an effective superexchange interaction $\sim J^2/U$, with the corresponding two-qubit Hamiltonian,
\begin{eqnarray}
H_\text{2-qubit} &=-& \frac{J^2}{U}\left[ \ket{01}\bra{01} +\ket{10}\bra{10} + \left(\ket{10}\bra{01} + \text{H.c.}\right) \right], \quad \ \
\end{eqnarray} 
where $\ket{s_1 s_2}=\ket{s_1}\otimes\ket{s_2} \ (s_{1,2}=0,1)$ are the computational basis for the two-qubit space. Since we have full control over the location of the vortices, we can tune our time-evolution operator to act as a $\sqrt{\text{SWAP}}$ gate up to some single-qubit $\sigma_z$ operations \cite{supplement}. This $\sqrt{\text{SWAP}}$  gate and previously introduced single-qubit rotations constitute a gate set for universal quantum computation \cite{divincenzo2000universal,PhysRevA.72.052323}. We stress again that this proximity-based scheme of two-qubit gate is only possible because the current system allows enhanced freedom to change the locations of Floquet vortex states. This is a big advantage that Floquet vortex state qubits have over other qubits based on solid-state systems such as quantum dots \cite{PhysRevA.57.120,kane1998silicon,PhysRevA.62.012306}.

While the state preparation in Floquet systems is a challenging problem in general, one may be able to prepare the desired Floquet state by using proper bosonic and fermionic reservoirs through dissipative engineering \cite{PhysRevB.90.195429, Esin2018quantized, Seetharam2019Steady}. Once the initialization method is established, the desired qubit state can be prepared by controlling the backgate voltage, similar to the initialization procedure in quantum-dot qubit systems.


\textit{Conclusions and outlook.--} To further elaborate the scheme for the quantum information processing, it would be interesting to study the possible measurement protocols for the OAM of the Floquet vortex states. One potential candidate for such protocol is through the measurement of optical Hall conductivity, which might have different responses on the states with different OAM. Also, since our system has multiple non-linearly-dispersed Floquet vortex states, the extension to the qudit system is a natural topic for future study. While we briefly examined the possibility of such vortex state as a qubit, there are a lot of unanswered questions such as the heating, decoherence, and sensing in this platform. While we treated the vortex state of a single electron, it would be interesting to study how the presence of electronic interactions can change the vortex state structure or even help to create exotic many-body states. Another interesting direction is to investigate lattices of optical vortices and other field patterns such as electromagnetic skyrmions \cite{tsesses2018optical}.


\textit{Acknowledgments.--} We thank stimulating discussions with H. Aoki. I.M. was supported by the Materials Sciences and Engineering Division, Basic Energy Sciences, Office of Science, U.S. Dept. of Energy. The work at Maryland was supported by ARL W911NF1920181, AFOSR MURI FA9550- 19-1-0399, FA95502010223, and Simons Foundation. 

\bibliography{biblio}

\clearpage
\onecolumngrid
\begin{center}
\textbf{\large Supplemental Material: Floquet vortex states induced by light carrying the orbital angular momentum}
\end{center}
\twocolumngrid
\setcounter{page}{1}

\appendix
\section{Rotating wave approximation (RWA)}
\setcounter{equation}{0}
\renewcommand{\theequation}{A\arabic{equation}}
As stated in the main text, we consider following model Hamiltonian $H_{0} = (vk_x,\pm vk_y,M)\cdot\sigma = \mathbf{D}_{\mathbf{k}}\cdot\sigma$ for our semiconductor.
We now consider the electromagnetic radiation $\mathcal{A}(\mathbf{r},t)$. Then the minimal coupling $\mathbf{k} \to \mathbf{k}+e\mathcal{A}(\mathbf{r},t)$ leads to the following time-dependent Hamiltonian,
\begin{eqnarray}
H(t) &=& H_{0} + ev\mathcal{A}(\mathbf{r},t) \cdot \sigma \nonumber\\
&=& H_{0} + V(t) = H_{0} + \left[\mathbf{V} e^{i\omega t} +\text{c.c.}\right]\cdot\sigma.
\end{eqnarray}
Then the projection operators to conduction and valence bands are
\begin{eqnarray}
& P_{c} = \int d^2\mathbf{k} P_{c, \mathbf{k}} = \int d^2\mathbf{k} (1 + \mathbf{d}_{\mathbf{k}})/2, \nonumber\\
& P_{ v} = \int d^2\mathbf{k} P_{v, \mathbf{k}} = \int d^2\mathbf{k} (1 - \mathbf{d}_{\mathbf{k}})/2,
\end{eqnarray}
where $\mathbf{d}_{\mathbf{k}} = \mathbf{D}_{\mathbf{k}}/|\mathbf{D}_{\mathbf{k}}|$. Considering the rotating frame $U(t) = P_{c} e^{-i\omega t} + P_{v} e^{i\omega t}$, the rotated Hamiltonian is
\begin{eqnarray}
H_{\text{rot}} &=& -i U^\dag(t) \partial_t U(t) + U^\dag(t) H(t) U(t) \nonumber\\
&=& \frac{\omega}{2}\left( P_{v} - P_{c} \right) + \mathbf{D}_{\mathbf{k}}\cdot\sigma + P_{c} V(t) P_{c} + P_{v} V(t) P_{v} \nonumber\\
&& + e^{i\omega t} P_{c} V(t) P_{v} + e^{-i\omega t} P_{v} V(t) P_{c}.
\end{eqnarray}
In the weak field regime $evA_\text{max} = \Omega_0 \ll \omega$, we can obtain RWA Hamiltonian by dropping fast oscillating terms from $H_\text{rot}$,
\begin{eqnarray}
H_{\text{RWA}} = \left( \mathbf{D}_{\mathbf{k}} - \frac{\omega}{2}\mathbf{d}_{\mathbf{k}} \right)\cdot\sigma + \mathcal{V}_{\mathbf{k}},
\end{eqnarray}
where
\begin{eqnarray}\label{Vk}
\mathcal{V}_{\mathbf{k}} &=& P_{c,\mathbf{k}} (\mathbf{V}^*\cdot\sigma) P_{v,\mathbf{k}} + P_{v,\mathbf{k}} (\mathbf{V}\cdot\sigma) P_{c,\mathbf{k}} \nonumber\\
&=& \left\lbrace \text{Re}\mathbf{V}\cdot\sigma 
+ i[\text{Im}\mathbf{V}\cdot\sigma, \mathbf{d}_{\mathbf{k}}\cdot\sigma] \right. \nonumber\\
&& \left. \ -(\mathbf{d}_{\mathbf{k}}\cdot\sigma)(\text{Re}\mathbf{V}\cdot\sigma)(\mathbf{d}_{\mathbf{k}}\cdot\sigma) \right\rbrace/2 \nonumber\\
&=& \frac{1}{2}\left[
\text{Re}\mathbf{V} + \left( \mathbf{d}_{\mathbf{k}} \times \text{Im}\mathbf{V} - \text{Im}\mathbf{V} \times \mathbf{d}_{\mathbf{k}} \right) \right. \nonumber\\
&& \ \left. -(\mathbf{d}_{\mathbf{k}} \cdot \text{Re}\mathbf{V})\mathbf{d}_{\mathbf{k}}
+(\mathbf{d}_{\mathbf{k}} \times \text{Re}\mathbf{V}) \times \mathbf{d}_{\mathbf{k}}
\right]\cdot\sigma \nonumber\\
&& +\frac{i}{2}\left[
\text{Im}\mathbf{V} \cdot \mathbf{d}_{\mathbf{k}}
-\mathbf{d}_{\mathbf{k}} \cdot \text{Im}\mathbf{V}
-(\mathbf{d}_{\mathbf{k}} \times \text{Re}\mathbf{V}) \cdot \mathbf{d}_{\mathbf{k}}
\right]. \qquad
\end{eqnarray}
For small detuning regime $\delta = \omega - 2M \ll \omega$, $\delta \simeq v^2 k_0^2 /M$ and $vk_0\ll M$. Then, for small momenta $|\mathbf{k}| = O(k_0)$,
\begin{eqnarray}
\mathbf{d}_{\mathbf{k}} &=& (d_{x,\mathbf{k}},d_{y,\mathbf{k}}, d_{z,\mathbf{k}})
= \frac{1}{\sqrt{M^2 + v^2\mathbf{k}^2}}(vk_x,vk_y, M) \nonumber\\
&=& \left( \frac{vk_x}{M}, \frac{vk_y}{M}, 1 - \frac{v^2\mathbf{k}^2}{2M^2} \right) 
+ O\left( \frac{v^3 k_0^3}{M^3} \right),
\end{eqnarray}
\begin{eqnarray}\label{Hz}
\left( \mathbf{D}_{\mathbf{k}} - \frac{\omega}{2}\mathbf{d}_{\mathbf{k}} \right)\cdot\sigma
&=& \left(1-\frac{\omega/2}{\sqrt{M^2 + v^2\mathbf{k}^2}}\right)(v k_x,vk_y, M)\cdot\sigma \nonumber\\
&=& \frac{v^2}{2M}(\mathbf{k}^2 - k_0^2)\sigma_z + O\left(\frac{v^3 k_0^3}{M^2}\right).
\end{eqnarray}
Now we consider a linearly polarized light carrying OAM, $\mathcal{A}(\mathbf{r},t) = \left[ A_0(r) e^{im\phi} e^{i\omega t} + \text{c.c.} \right]\mathbf{\hat{x}}$. With this, $V_y = 0$, and from \eqnref{Vk},
\begin{eqnarray}
\mathcal{V}_{\mathbf{k}} &=& \frac{1}{2}\left[
\left( \text{Re}V_x + d_{z,\mathbf{k}} \text{Re}V_x d_{z,\mathbf{k}} \right.\right.\nonumber\\
&&\qquad \left.-d_{x,\mathbf{k}} \text{Re}V_x d_{x,\mathbf{k}} + d_{y,\mathbf{k}}\text{Re}V_x d_{y,\mathbf{k}} \right)\sigma_x \nonumber\\
&& \quad + \left( d_{z,\mathbf{k}} \text{Im}V_x + \text{Im}V_x d_{z,\mathbf{k}} \right. \nonumber\\
&& \qquad \left. - d_{x,\mathbf{k}} \text{Re}V_x d_{y,\mathbf{k}}
- d_{y,\mathbf{k}} \text{Re}V_x d_{x,\mathbf{k}} \right)\sigma_y \nonumber\\
&& \quad + \left(
- d_{y,\mathbf{k}} \text{Im}V_x -d_{x,\mathbf{k}} \text{Re}V_x d_{z,\mathbf{k}} \right.\nonumber\\
&& \qquad \left. \left. -d_{z,\mathbf{k}} \text{Re}V_x d_{x,\mathbf{k}} \right)\sigma_z \right] \nonumber\\
&& +\frac{i}{2}\left(
\text{Im}V_x d_{x,\mathbf{k}} - d_{x,\mathbf{k}} \text{Im}V_x \right. \nonumber\\
&& \qquad \left. - d_{z,\mathbf{k}} \text{Re}V_x d_{y,\mathbf{k}}
+ d_{y,\mathbf{k}} \text{Re}V_x d_{z,\mathbf{k}} \right) \nonumber\\
&=& \text{Re}V_x \sigma_x + \text{Im}V_x \sigma_y + O\left(evA_\text{max}\frac{vk_0}{M}\right).
\end{eqnarray}
Therefore, with further assumption of weak field $\Omega_0 \ll \sqrt{\delta M}$, the RWA Hamiltonian becomes
\begin{eqnarray}
H_{\text{RWA}} &=& \frac{v^2}{2M}\left(\mathbf{k}^2 - k_0^2\right)\sigma_z + \left[evA_0(r) e^{-im\phi} \sigma_+ + \text{H.c.}\right] \nonumber\\
&& + O\left(evA_\text{max}\frac{vk_0}{M}\right) \nonumber\\
&=& \frac{\delta}{2}\left(\frac{\mathbf{k}^2}{k_0^2} - 1\right)\sigma_z + \left[ \Omega(r) e^{-im\phi} \sigma_+ + \text{H.c.} \right] \nonumber\\
&& + O\left(\Omega_0\sqrt{\frac{\delta}{M}}\right),
\end{eqnarray}
so we derived the RWA Hamiltonian in \eqnref{H_rwa}.

Due to the OAM of the light, the RWA Hamiltonian $H_\text{RWA}$ and the static semiconductor Hamiltonian $H_0$ have different symmetries. While $H_0$ commutes with electronic OAM $-i\partial_\phi$, $H_\text{RWA}$ commutes with pseudo-OAM $\hat{l} = -i\partial_\phi + (m/2)\sigma_z$. To demonstrate this, we use $[-i\partial_\phi,k_x]=ik_y$ and $[-i\partial_\phi,k_y]=-ik_x$. These yield $[-i\partial_\phi,k_x\pm ik_u]=\pm(k_x\pm ik_y)$ and $[-i\partial_\phi,\mathbf{k}^2]=0$, therefore
\begin{eqnarray}
\left[-i\partial_\phi, H_{\text{RWA}}\right] &=& -m\left(\Omega(r) e^{-im\phi }\sigma_+ -\text{H.c.} \right), \nonumber\\
\left[\sigma_z, H_{\text{RWA}}\right] &=& 2\left( \Omega(r) e^{-im\phi} \sigma_+ -\text{H.c.} \right),
\end{eqnarray}
so $[-i\partial_\phi + (m/2)\sigma_z, H_{\text{RWA}}]=0$. Since $l$ is a good quantum number, the wave functions for each $l$ have the form of
\begin{eqnarray}\label{psi_l}
\psi_{n,l}(\mathbf{r}) = \left( e^{i(l-m/2)\phi} u_{n,+}(r), e^{i(l+m/2)\phi} u_{n,-}(r) \right)^T,\quad
\end{eqnarray}
where $n$ is the branch index. With this, $H_{\text{RWA}}$ leads to following eigenvalue problem for each $l$,
\begin{eqnarray}
E_{n,l} u_{n,l,\pm}(r) &=& \mp \frac{\delta^2}{2k_0^2}
\left( \partial_r^2 + \frac{1}{r}\partial_r - \frac{(l\mp m/2)^2}{r^2} + k_0^2 \right) u_{n,l,\pm}(r) \nonumber\\
&& + \Omega(r) u_{n,l,\mp}(r).
\end{eqnarray}
By observing this Hamiltonian, one can see this Hamiltonian preserves the particle-hole symmetry $\psi_{n,-l}(\mathbf{r})=i\sigma_y \psi_{|m|+1-n,l}^*(\mathbf{r})$ and $E_{n,-l}=-E_{|m|+1-n,l}$. Here the branch index $n$ should alter to $|m|+1-n$ as $l$ changes to $-l$.

\section{Number of Floquet vortex states branches}
\setcounter{equation}{0}
\renewcommand{\theequation}{B\arabic{equation}}
Since $H_\text{RWA}(\mathbf{k})$ in \eqnref{H_rwa} is particle-hole symmetric and gapped except the vortex core, the intragap modes develop around the vortex core are expected to cross the zero energy, if any exists. We may use the semiclassical approach introduced in Ref. \cite{volovik1993vortex} to investigate the number of such intragap modes. Let us consider the Hamiltonian in the classical regime, $H_\text{RWA}\to \mathbf{H} \cdot \sigma$, where the momentum and the position commute each other. This semiclassical treatment is justified as long as $k_0\xi \gg 1$. Here, the vector $\mathbf{H}=\mathbf{H}(k,r,\phi)$ resides on the 3D parameter space $(k,r,\phi)$. Now such Hamiltonian yields energy $E^2(k,r,\phi) = |\mathbf{H}(k,r,\phi)|^2 = \delta^2(k^2/k_0^2-1)^2/4 + \Omega(r)^2$ and $E=0$ is achieved at $k=k_0$ and $r=0$. To consider the surface surrounds this zero point, let us consider the surface $|E|=\Delta E$ for small energy $\Delta E$. Such surface would be located in the vicinity of that zero point, so we can write $k=k_0+\Delta k$ and $r=\Delta r$. To the leading order, This surface can be written as $\Delta E^2 = (\delta /k_0)^2\Delta k^2 +\Omega(\Delta r)^2$. Without loss of generality, we can regard $\Omega(\Delta r)=\lambda \Delta r$. Now the surface $|E|=\Delta E$ becomes an ellipsoid and can be parameterized by the polar angle $\theta$ and the azimuthal angle $\phi$: $\Delta k = (k_0\Delta E /\delta)\cos\theta$, $\Delta r = (\Delta E/\lambda)\sin\theta$, $\Delta x=\Delta r \cos\phi$, $\Delta y=\Delta r \sin\phi$. Then the skyrmion number of $\mathbf{H}$ on this ellipsoid is equal to the number of branches that passes the zero energy in the intragap spectrum. Since the skyrmion number is a topological invariant, we did not lose the generality even if the actual behavior of $\Omega(r)$ for small $r$ is not linear. For the current parameterization,
\begin{eqnarray}
&&\left.\mathbf{H}\right|_{|E|=\Delta E,\phi_k} \nonumber\\
&&= \left.\Omega(r)\left[\cos(m\phi)\mathbf{\hat{x}} +\sin(m\phi)\mathbf{\hat{y}}\right] + \frac{\delta}{2}\left(\frac{k^2}{k_0^2}-1\right)\mathbf{\hat{z}} \right|_{|E|=\Delta E,\phi_k} \nonumber\\
&& = \Delta E \left[ \sin\theta \left( \cos(m\phi)\mathbf{\hat{x}} +\cos(m\phi)\mathbf{\hat{y}} \right) +\cos\theta\mathbf{\hat{z}} \right]\nonumber\\
&& = \Delta E \ \mathbf{\hat{H}}(\theta,\phi),
\end{eqnarray}
and now the skyrmion number is calculated as
\begin{eqnarray}
N_\text{mid} &=& \frac{1}{4\pi}\int_0^{2\pi}d\phi \int_0^\pi d\theta
\left(\frac{\partial \mathbf{\hat{H}}}{\partial \theta}\times\frac{\partial \mathbf{\hat{H}}}{\partial \phi}\right)\cdot \mathbf{\hat{H}} \nonumber\\
&=& \frac{1}{4\pi}\int_0^{2\pi}d\phi \int_0^\pi d\theta \ m\sin\theta = m.
\end{eqnarray}
Note that the number of intragap branches $N_\text{mid}$ is solely determined by the winding number of the applied field, regardless of the winding number along the momentum direction. Yet, the presence of intragap branches crossing the zero energy does not guarantee the existence of the exact zero mode, since the mini gap can develop within each branch in the process of quantization. For further analysis, a fully quantum mechanical approach is required.

\section{Estimation of energy separations in large optical vortex regime}
\setcounter{equation}{0}
\renewcommand{\theequation}{C\arabic{equation}}

Following the formalism in Ref. \cite{PhysRevLett.119.067003}, we find the energy separations between the Floquet vortex states and the intragap state branches, respectively,
\begin{eqnarray}
\omega_0 &=& \frac{\int_0^\infty \frac{\Omega(r)}{r} e^{-(2k_0/\delta)\int_0^r \Omega(r') dr'} dr}{k_0\int_0^\infty e^{-(2k_0/\delta)\int_0^r \Omega(r') dr'} dr}, \nonumber\\
\tilde{\omega}_0 &=& \frac{\delta (\pi/2)}{k_0\int_0^\infty e^{-(2k_0/\delta)\int_0^r \Omega(r') dr'} dr},
\end{eqnarray}
for low energy, low pseudo-OAM, and large optical vortex regime, as explained in \eqnref{lin_disp}.

In this appendix, we demonstrate how these energy separations depend on radiation parameters such as $\Omega_0$, $\delta$, $\omega$ as well as the radial profile of the applied light beam. For this, we estimate $\omega_0$ and $\tilde{\omega}_0$ for variants of radial beam profile. Specifically, we consider the radial profile
\begin{eqnarray}
\Omega(r) = \left\lbrace
\begin{array}{cc}
\Omega_0 (r/\xi)^q & \text{ for } r\le\xi \\
\Omega_0  & \text{ for } r>\xi
\end{array}
\right.,
\quad q \ge 1.
\end{eqnarray}
With this, we define $\mathcal{F}(r) \equiv \exp\left[-(2k_0/\delta)\int_0^r \Omega(r') dr' \right]$ and it becomes
\begin{eqnarray}
\mathcal{F}(r) = \left\lbrace
\begin{array}{cc}
\exp\left[ -\frac{2 k_\delta \xi}{q+1}\left(\frac{r}{\xi}\right)^{q+1} \right] & \text{ for } r\le\xi \\
\exp\left[-2 k_\delta \left(r-\frac{q}{q+1}\xi\right)\right] &
\text{ for } r > \xi
\end{array}
\right..\quad
\end{eqnarray}

With $k_\delta \xi \gg 1$, this $\mathcal{F}(r)$ can be roughly estimated by a step function $\theta(x) \equiv [\text{sgn}(x)+1]/2$,
\begin{eqnarray}
\mathcal{F}(r) \simeq
\theta\left(r_\text{cut} -r\right), \
r_\text{cut} = O\left(\xi \left( \frac{q+1}{2 k_\delta \xi} \right)^{1/(q+1)}\right).\qquad
\end{eqnarray}
With this,
\begin{eqnarray}
&& \int_0^\infty \frac{\Omega(r)}{r} \mathcal{F}(r) dr \simeq \int_0^{r_\text{cut}}\Omega_0 \frac{r^{q-1}}{\xi^q} dr
= \frac{\Omega_0 r_\text{cut}^{q}}{q\xi^q} \nonumber\\
&& \qquad\qquad\qquad = O\left(\Omega_0[k_\delta \xi]^{-q/(q+1)}\right), \nonumber\\
&& \int_0^\infty \mathcal{F}(r) dr
\simeq  r_\text{cut} = O\left(\xi(k_\delta \xi)^{-1/(q+1)}\right),
\end{eqnarray}
then
\begin{eqnarray}
\omega_0 &=& \frac{\int_0^\infty r^{-1} \Omega(r) \mathcal{F}(r) dr}{k_0 \int_0^\infty \mathcal{F}(r) dr} \nonumber\\
&\simeq & O\left(
\Omega_0 (k_0\xi)^{-1} (k_\delta \xi)^{-(q-1)/(q+1)}
\right) \nonumber\\
\tilde{\omega}_0 &=& \frac{\delta (\pi/2)}{k_0 \int_0^\infty \mathcal{F}(r) dr} \nonumber\\
&\simeq & O\left(
\delta (k_0\xi)^{-1} (k_\delta \xi)^{1/(q+1)}
\right).
\end{eqnarray}
As seen in this estimation, energy separations $\omega_0$ and $\tilde{\omega}_0$ depend not only on radiation parameters like $\Omega_0$, $\delta$, $\omega$, but also on parameters related to the size ($\xi$) and shape ($q$) of the radial profile of the beam.

From these results, we can further estimate the number of vortex modes in a branch as
\begin{eqnarray}
2\Omega_0 /\omega_0 = O\left( k_0 \xi (k_\delta \xi)^{(q-1)/(q+1)} \right).
\end{eqnarray}
Also, we can estimate the angular momentum difference between branches as
\begin{eqnarray}
\tilde{\omega}_0 / \omega_0 = O\left(
\frac{k_0}{k_\delta}
(k_\delta \xi)^{q/(q+1)}
\right).
\end{eqnarray}
With $q\ge 1$, the lower bound of these estimations are given as $2\Omega_0 /\omega_0 = O\left( k_0 \xi\right)$ and $\tilde{\omega}_0 / \omega_0 = O\left(k_0 k_\delta^{-1} \sqrt{k_\delta \xi}\right)$.

\section{Illumination of circularly polarized light}
\setcounter{equation}{0}
\renewcommand{\theequation}{D\arabic{equation}}

The hybridization gap for the bulk part of systems with linearly polarized light is in the order of $\Omega_0$. For the most of systems with different beam polarization, it is still true and therefore results in similar RWA Hamiltonian with \eqnref{H_rwa}. However, the situation is different for circularly polarized light. As explained in Ref. \cite{lindner2011floquet}, a semiconductor valley with valley Hamiltonian $H_{0,\pm} = (vk_x,\pm vk_y,M)$ becomes a Floquet Chern insulator when illuminated by circularly polarized light $\mathcal{A}_\pm(\mathbf{r},t) = A(\mathbf{r})(\mathbf{\hat{x}}\pm i\mathbf{\hat{y}})e^{i\omega t} + \text{c.c.}$. In such Floquet Chern insulator, the size of hybridization gap is in the order of $\delta \Omega_0/M$, instead of $\Omega_0$. In this appendix, we derive the RWA Hamiltonian for the light carrying OAM with this circular polarization. Then we calculate the wavefunctions and dispersion of Floquet vortex states given by that Hamiltonian. For simplicity, we only consider the valley Hamiltonian $H_0 = H_{0,+}$ and the field $\mathcal{A}(\mathbf{r},t) = \mathcal{A}_+(\mathbf{r},t)$ from now on.

The RWA Hamiltonian derived in appendix A is valid regardless of $\mathcal{A}(\mathbf{r},t)$ up to \eqnref{Hz}. By using $\mathcal{A}(\mathbf{r},t) = A(\mathbf{r})(\mathbf{\hat{x}}+ i\mathbf{\hat{y}})e^{i\omega t} + \text{c.c.}$, we have $V_y = i V_x$. This yields
\begin{eqnarray}
\mathcal{V}_{\mathbf{k}} &=& \frac{1}{2}\left[
\text{Re}\mathbf{V} + \left( \mathbf{d}_{\mathbf{k}} \times \text{Im}\mathbf{V} - \text{Im}\mathbf{V} \times \mathbf{d}_{\mathbf{k}} \right) \right. \nonumber\\
&& \ \left. -(\mathbf{d}_{\mathbf{k}} \cdot \text{Re}\mathbf{V})\mathbf{d}_{\mathbf{k}}
+(\mathbf{d}_{\mathbf{k}} \times \text{Re}\mathbf{V}) \times \mathbf{d}_{\mathbf{k}}
\right]\cdot\sigma \nonumber\\
&& +\frac{i}{2}\left[
\text{Im}\mathbf{V} \cdot \mathbf{d}_{\mathbf{k}}
-\mathbf{d}_{\mathbf{k}} \cdot \text{Im}\mathbf{V}
-(\mathbf{d}_{\mathbf{k}} \times \text{Re}\mathbf{V}) \cdot \mathbf{d}_{\mathbf{k}}
\right] \nonumber\\
&=& (1- d_{z,\mathbf{k}}) (\text{Re} V_{x} \sigma_x - \text{Im} V_{x} \sigma_y) (1- d_{z,\mathbf{k}})/2 \nonumber\\
&& -\frac{1}{2} \left[(d_{x,\mathbf{k}}-id_{y,\mathbf{k}})(\text{Re}V_{x}- i \text{Im}V_{x})(d_{x,\mathbf{k}}- id_{y,\mathbf{k}})\sigma_+ \right.\nonumber\\
&&\qquad \left. + \text{H.c.}\right] + O\left(\frac{evA_\text{max} v^3 k_0^3}{M^3}\right).
\end{eqnarray}
Then the RWA Hamiltonian becomes
\begin{eqnarray}\label{Hrwa_cp}
H_{\text{RWA}} &=& -\frac{ev^3}{2M^2}\left[ (k_x + ik_y)A_0(r) e^{im\phi} (k_x + ik_y) \sigma_- + \text{H.c.} \right] \nonumber\\
&& + \frac{v^2}{2M}(\mathbf{k}^2 - k_0^2)\sigma_z + O\left(\frac{v^3 k_0^3}{M^2}\right) \nonumber\\
&=& -\frac{\delta}{2M}\left[ \frac{(k_x + ik_y)}{k_0}\Omega(r) e^{im\phi} \frac{(k_x + ik_y)}{k_0} \sigma_- + \text{H.c.} \right] \nonumber\\
&& + \frac{\delta}{2}\left(\frac{\mathbf{k}^2}{k_0^2} - 1\right)\sigma_z + O\left( \delta \sqrt{\frac{\delta}{M}} \right).
\end{eqnarray}
In the bulk far from $r=0$, this system becomes a Floquet Chern insulator and therefore hosts edge states in the middle of hybridization gap. These states are localized at the boundary of the sample and has nothing to do with the OAM of the beam. We aim to find fully quantum mechanical solution for intragap states localized around the optical vortex. For this, we use a similar method used in Ref. \cite{caroli1964bound,PhysRevLett.114.195301,PhysRevB.93.174505,PhysRevLett.119.067003}. Note that, due to scale change, we redefine $k_\delta = k_0\Omega_0 /M$ for this section.

For the simplicity of discussion, we normalize the RWA Hamiltonian as $h
=(M/v^2)H_\text{RWA}$. We first demonstrate that $h$ commutes with pseudo-OAM $\hat{l} = -i\partial_\phi+(m/2+1)\sigma_z$. Note that the pseudo-OAM operator here differs from the pseudo-OAM operator for the systems with non-circularly polarized light by an extra term of $\sigma_z$. Similar to the linear polarization case, we use $[-i\partial_\phi,k_x]=ik_y$, $[-i\partial_\phi,k_y]=-ik_x$, $[-i\partial_\phi,k_x\pm ik_u]=\pm(k_x\pm ik_y)$, and $[-i\partial_\phi,\mathbf{k}^2]=0$, therefore
\begin{eqnarray}
&&[-i\partial_\phi,h] \\
&& = -(m+2)\frac{\Omega(r)}{2M}\left[(k_x-ik_y)e^{-im\phi}(k_x-ik_y)\sigma_+ -\text{H.c.} \right], \nonumber\\
&& [\sigma_z,h] =\frac{\Omega(r)}{M}\left[(k_x-ik_y)e^{-im\phi}(k_x-ik_y)\sigma_+ -\text{H.c.} \right], \nonumber
\end{eqnarray}
so we eventually have $[-i\partial_\phi+(m/2+1)\sigma_z,h]=0$. Therefore, $l$ is a conserved quantity and we can block-diagonalize $h$ along this $l$. Within the block for $l$, wavefunctions can be written as in \eqnref{psi_angular},
\begin{eqnarray}
\psi_l(\mathbf{r})=\left( e^{il_+\phi}u_{+}(r) , e^{il_-\phi}u_{-}(r) \right)^T, \qquad
\end{eqnarray}
where $l_\pm= l\mp(m/2+1)$. The eigenstates satisfy
\begin{eqnarray}\label{heff}
\epsilon u_+(r) &=& -\frac{1}{2}\left(\partial_r^2 +\frac{1}{r}\partial_r -\frac{l_+^2}{r^2} + k_0^2 \right)u_+(r) \nonumber\\
&&+\frac{\Omega(r)}{2M}\left( \partial_r^2 + \frac{2l+1}{r}\partial_r + \frac{l_+ l_-}{r^2} \right)u_-(r) \nonumber\\
&& +\frac{\Omega'(r)}{2M}\left( \partial_r + \frac{l_-}{r} \right) u_-(r), \nonumber\\
\epsilon u_-(r) &=& \frac{1}{2}\left(\partial_r^2 +\frac{1}{r}\partial_r -\frac{l_-^2}{r^2} + k_0^2 \right)u_-(r) \nonumber\\
&&+\frac{\Omega(r)}{2M}\left( \partial_r^2 -\frac{2l-1}{r}\partial_r + \frac{l_+ l_-}{r^2} \right)u_+(r) \nonumber\\
&& +\frac{\Omega'(r)}{2M}\left( \partial_r - \frac{l_+}{r} \right) u_+(r).
\end{eqnarray}
As in the system with linearly polarized light, this RWA Hamiltonian preserves the particle-hole symmetry. By replacing $l$ by $-l$ in this equation, $l_\pm \to - l_\mp$, so one can readily show that $\psi_{-l}(\mathbf{r})=i\sigma_y \psi_l^*(\mathbf{r})$ with $\left.\epsilon\right|_{-l} = -\left.\epsilon\right|_{l}$. Equivalent to \eqnref{heff},
\begin{eqnarray}\label{heff2}
\left( \epsilon +\frac{\beta}{2r^2} \right)u_+(r)
&&= -\frac{1}{2}\left(\partial_r^2 +\frac{1}{r}\partial_r -\frac{\alpha^2}{r^2} + k_0^2 \right)u_+(r) \nonumber\\
&&+\frac{\Omega(r)}{2M}\left( \partial_r^2 + \frac{2l+1}{r}\partial_r + \frac{l_+ l_-}{r^2} \right)u_-(r) \nonumber\\
&& +\frac{\Omega'(r)}{2M}\left( \partial_r + \frac{l_-}{r} \right) u_-(r), \nonumber\\
\left( \epsilon +\frac{\beta}{2r^2} \right)u_-(r) &&= \frac{1}{2}\left(\partial_r^2 +\frac{1}{r}\partial_r -\frac{\alpha^2}{r^2} + k_0^2 \right)u_-(r) \nonumber\\
&&+\frac{\Omega(r)}{2M}\left( \partial_r^2 -\frac{2l-1}{r}\partial_r + \frac{l_+ l_-}{r^2} \right)u_+(r) \nonumber\\
&& +\frac{\Omega'(r)}{2M}\left( \partial_r - \frac{l_+}{r} \right) u_+(r),
\end{eqnarray}
where $\alpha=\sqrt{l^2+(m/2+1)^2}$ and $\beta=l(m+2)$. While it is difficult to find the generic solution for this equation, we can find the low-energy solution for the regime $l^2/k_0\ll k_\delta^{-1} \ll \xi$. Let us consider a radius $r^*$ such that $l^2/k_0 \ll r^* \ll k_\delta^{-1}$. For $r\ll r^*$, $\Omega(r)\to 0$ and therefore we can decouple $u_+(r)$ and $u_-(r)$ in \eqnref{heff},
\begin{eqnarray}
\left( \partial_r^2+\frac{1}{r}\partial_r-\frac{l_\pm^2}{r^2} + k_0^2\pm 2\epsilon \right)u_\pm(r)=0,
\end{eqnarray}
which yields the solution
\begin{eqnarray}\label{u_smallr}
u_\pm(r) = C_\pm J_{l\mp(m/2+1)}\left(\sqrt{(k_0^2\pm 2\epsilon)}r\right)
\end{eqnarray}
where $J_\nu(r)$ is the Bessel function of the first kind. The Bessel function of the second kind can be ruled out since the solution should be finite at $r=0$. In the low-energy theory, $\epsilon\ll k_0^2$, we can write $\sqrt{k_0^2\pm 2\epsilon}\simeq k_0\pm p$ where $p=\epsilon/k_0\ll k_0$.

For $r\gg r^*$, we take the ansatz
\begin{eqnarray}
u_\pm(r) = f_\pm(r) H_\alpha^{(1)}(k_0 r) + g_\pm(r) H_\alpha^{(2)}(k_0 r)
\end{eqnarray}
where $H_\nu^{(1)}(x), H_\nu^{(2)}(x)$ are the Hankel functions of the first kind and the second kind. Let us deal with the solutions for $f_\pm(r)$ first. Let us denote $H_\alpha^{(1)}(x)=H(x)$ for short. Denoting that $(\partial_r^2 +r^{-1}\partial_r -\alpha^2/r^2+k_0^2)H(k_0 r)=0$, \eqnref{heff2} can be written as
\begin{eqnarray}\label{hankel}
&& \left(\epsilon +\frac{\beta}{2r^2}\right)f_+ H = -\frac{1}{2}\left( f''_+ H + 2f'_+ H' + \frac{f'_+ H}{r} \right) \nonumber\\
&& +\frac{\Omega}{2M}\left( f''_- H + 2f'_- H' + f_- H''
+\frac{2l+1}{r}(f'_- H + f_- H') \right. \nonumber\\
&& \qquad\left. +\frac{l_+ l_-}{r^2}f_- H \right)
+\frac{\Omega'}{2M}\left( f'_- H + f_- H' +\frac{l_-}{r} \right), \nonumber\\
&& \left(\epsilon +\frac{\beta}{2r^2}\right)f_- H = \frac{1}{2}\left( f''_- H + 2f'_- H' + \frac{f'_- H}{r} \right) \nonumber\\
&& +\frac{\Omega}{2M}\left( f''_+ H + 2f'_+ H' + f_+ H''
-\frac{2l-1}{r}(f'_+ H + f_+ H') \right. \nonumber\\
&& \qquad\left. +\frac{l_+ l_-}{r^2}f_+ H \right)
+\frac{\Omega'}{2M}\left( f'_+ H + f_+ H' -\frac{l_+}{r} \right).
\end{eqnarray}
To simplify these equations, we estimate and compare the magnitude of different terms in these equations around $r=k_\delta^{-1}$. For this, we take the ansatz $f_\pm(r)=f_{\pm,(0)}(r)\exp[i\eta_\pm(r)]$ where $f'_{\pm,(0)}/f_{\pm,(0)}=O(k_\delta)$, $\eta_\pm=O(k_\delta/k_0)$, and $\eta'_\pm/\eta_\pm=O(k_\delta)$ around $r=(k_\delta)^{-1}$. We further restrict the eigenenergy to be $\epsilon=O\left(k_\delta^2\right)$. Assuming $|f_{+,(0)}/f_{-,(0)}|=O(1)$ and noting that $\partial_r H_\alpha^{(1)}(k_0 r)\simeq ik_0 H_\alpha^{(1)}(k_0 r)$ for $k_0 r \gg l^2$, the lowest order equations of \eqnref{hankel} become
\begin{eqnarray}
&& O(k_0 k_\delta):  \mp i k_0 f'_{\pm,(0)}- k_0^2 \Omega (2M)^{-1} f_{\mp,(0)}=0, \nonumber\\
&& O\left(k_\delta^2\right): \ 
\left(\epsilon +\frac{\beta}{2r^2}\right)f_{\pm,(0)} = k_0 f'_{\pm,(0)}\eta_\pm + k_0 f_{\pm,(0)} \eta'_\pm \nonumber\\
&&\quad \mp\frac{f'_{\pm,(0)}}{2r} \pm i\frac{k_0^2 \Omega}{2M} f_{\mp,(0)}\eta_\mp \pm i\frac{k_0\Omega}{2M r}(2l\pm 1)f_{\mp,(0)}. \ \qquad
\end{eqnarray}
By solving the equations of the order of $O(k_0 k_\delta)$, we get $f_{+,(0)}=B\exp\left(-\frac{1}{2}(k_0/M)\int_0^r\Omega(r')dr'\right)=-if_{-,(0)}$. This solution indeed satisfies the supposition $f'_{\pm,(0)}/f_{\pm,(0)}=O(k_\delta)$. Then the equations of the order of $O\left(k_\delta^2\right)$ become
\begin{eqnarray}
k_0 \eta'_\pm -\frac{k_0^2 \Omega}{2M}(\eta_+ +\eta_-) = \epsilon +\frac{\beta}{2r^2} +\frac{k_0\Omega}{ M r}\left(l\pm\frac{1}{4}\right),\qquad
\end{eqnarray}
or equivalently,
\begin{eqnarray}
&& k_0\partial_r(\eta_+ +\eta_-)-\frac{k_0^2\Omega}{M}(\eta_+ +\eta_-)=2\epsilon +\frac{\beta}{r^2}+2l\frac{k_0\Omega}{M r}, \nonumber\\
&& k_0\partial_r(\eta_+ -\eta_-)=\frac{k_0\Omega}{2M r}.
\end{eqnarray}
The solutions of these equations can be found as
\begin{eqnarray}\label{eta}
&&\eta_+(r)+\eta_-(r)=-\frac{2}{k_0} e^{\frac{k_0}{M}\int_0^r\Omega(r')dr'}\nonumber\\
&&\qquad \times\int_r^\infty dr'\left(\epsilon +\frac{\beta}{2r'^2}+l\frac{k_0\Omega(r')}{M r'}\right) e^{-\frac{k_0}{M}\int_0^{r'}\Omega(r'')dr''}, \nonumber\\
&&\eta_+(r)-\eta_-(r)=\int_0^r \frac{\Omega(r')}{2M r'}dr'.
\end{eqnarray}
We have $(k_0/M)\int_0^{1/k_\delta}\Omega(r)dr=O(1)$, $\epsilon+\beta/(2r^2)+lk_0\Omega(r)/(M r)\le O\left(k_\delta^2\right)$ for $r\ge O(k_\delta^{-1})$, and $\lim_{r\to 0}\Omega(r)/r<\infty$, so the suppositions $\eta_\pm=O(k_\delta/k_0)$ and $\eta'_\pm/\eta_\pm=O(k_\delta)$ are justified around $r=(k_\delta)^{-1}$. One might worry that $\eta_+(r)-\eta_-(r)$ diverges as $r\to\infty$, but $\eta_+(r)-\eta_-(r)$ is bounded to $O(k_\delta/k_0)$ as long as $r\le O\left(k_\delta^{-1}\right)$ and the wavefunction vanishes for $r \gg k_\delta^{-1}$ due to the behaviors of $f_{\pm,(0)}(r)$, so the solutions become consistent.

\begin{figure}[t]
\centering
\includegraphics[width=1.0\linewidth]{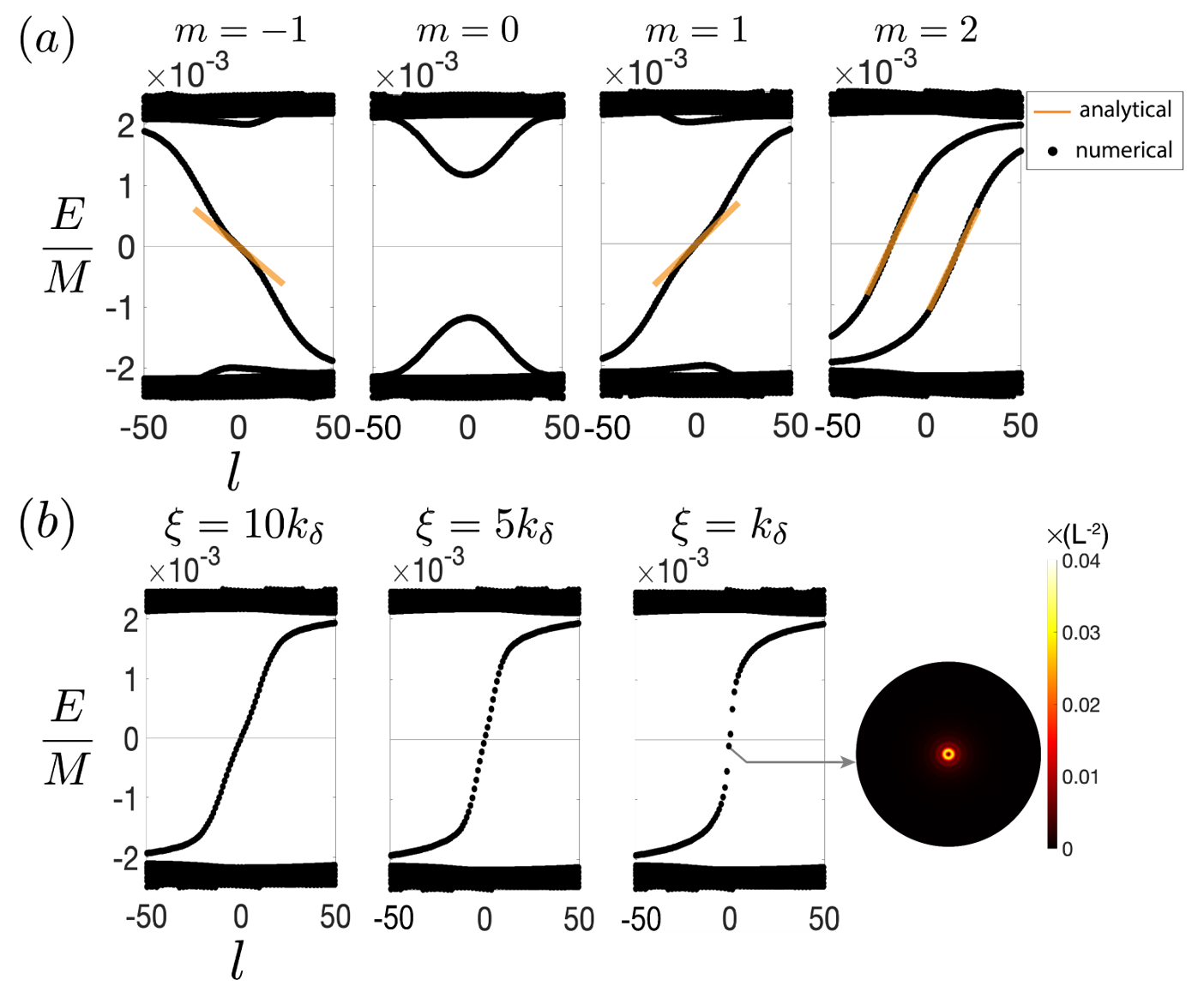}\subfigure{\label{F5a}}\subfigure{\label{F5b}}
\caption{(a) Numerically calculated energy spectra as a function of pseudo-OAM $l$. We use $\omega=2.05M$, $A_\text{max}=0.09M (ev)^{-1}$, and $A_0(r)=A_\text{max}\left[1-\exp\lbrace-r^2/(2\xi^2)\rbrace\right]$, $\xi=20k_{\delta}$, and the disk sample of radius $25\xi$. The numerical energy dispersions agree with the analytically expected spectra in \label{cp_disp} that includes the number of intragap state branches and the slope of the linear dispersion for small $|E_l|$ and $l$. (b) Demonstration of dispersions' dependence on optical vortex size $\xi$ for $m=1$ with identical parameters with (a) except $\xi$ and the disk size that the latter is fixed on $500k_{\delta}$. The linear region of the dispersion shrinks and the energy separation between subsequent states increases as $\xi$ decreases. On the right-hand side, the electronic density profile of the vortex state for $m=1$ just below the zero of the energy is illustrated.}\label{F5}
\end{figure}

We can also obtain the solutions for $g_\pm(r)$ by taking the complex conjugate on \eqnref{hankel} since $\partial_r H_\alpha^{(2)}(k_0 r)\simeq -ik_0H_\alpha^{(2)}(k_0 r)$, therefore $g_\pm(r) = f^*_\pm(r)$. Finally, we can write down $u_\pm(r)$ for $r\gg r^*$ as
\begin{eqnarray}\label{u_larger}
u_\pm(r) &=& i^{(1\mp 1)/2}Be^{-\frac{1}{2}(k_0/M)\int_0^r \Omega(r')dr'} \\
&& \times\left( e^{\pm i[\eta_\pm(r)\pm\kappa]} H_\alpha^{(1)}(k_0 r)
\pm e^{\mp i[\eta_\pm(r)\pm\kappa]} H_\alpha^{(2)}(k_0 r) \right) \nonumber
\end{eqnarray}
for some relative phase $\kappa$. Now let us match the solutions in \eqnref{u_smallr} and \eqnref{u_larger} at $r=r^*$. For this, with $k_0 r^* \gg l^2$, we can use the asymptotic forms of Bessel functions,
\begin{eqnarray}
&& J_\nu(x)\simeq\sqrt{\frac{2}{\pi x}}\cos\left(x-\frac{2\nu+1}{4}\pi+\frac{4\nu^2-1}{8x}\right), \\
&& H_\nu^{(1),(2)}(x)\simeq\sqrt{\frac{2}{\pi x}}\exp\left[\pm i\left(x-\frac{2\nu+1}{4}\pi+\frac{4\nu^2-1}{8x}\right)\right], \nonumber
\end{eqnarray}
for $\nu=O(l)$. Now from \eqnref{u_smallr},
\begin{eqnarray}\label{u_smallr_asymp}
&& u_\pm(r^*) \simeq C_\pm \sqrt{\frac{2}{\pi(k_0\pm p)r^*}} \\
&& \times\cos\left((k_0\pm p)r^*-\frac{2l\mp m\mp 2 +1}{4}\pi+\frac{(2l\mp m\mp 2)^2-1}{8(k_0\pm p)r^*}\right). \nonumber
\end{eqnarray}
By matching the constant factor in \eqnref{u_larger} as $B=C_+$, we have
\begin{eqnarray}\label{u_larger_asymp}
u_\pm(r^*) &\simeq & C_+ \sqrt{\frac{2}{\pi k_0 r^*}}e^{-\frac{1}{2}(k_0/M)\int_0^{r^*}\Omega(r) dr} \nonumber\\
&& \times\cos\left( \pm\eta_\pm(r^*)+\kappa +k_0 r^* -\frac{2\alpha+1}{4}\pi \right.\nonumber\\
&&\qquad \ \left. + \frac{4\alpha^2-1}{8k_0 r^*} +\frac{1\mp 1}{2}\left(n-\frac{1}{2}\right)\pi \right) \qquad
\end{eqnarray}
where $n$ is odd integer. Now by comparing \eqnref{u_smallr_asymp} and \eqnref{u_larger_asymp}, we have
\begin{eqnarray}\label{phase_match}
&&\pm\eta_\pm(r^*) +\kappa \mp pr^* +\frac{2l-2\alpha\mp m \mp 2}{4}\pi \pm \frac{\beta}{2k_0 r^*} \nonumber\\
&& +\frac{1\mp 1}{2}\left(n-\frac{1}{2}\right)\pi = O\left(\frac{p}{k_0^2 r^*}\right).
\end{eqnarray}
Here, we now let $n$ be any integer by using the freedom to choose the sign of $C_+/C_-$. In fact, this $n$ serves as the branch index, so we put this branch index for each state from now on. That is, $\psi_l \ to \psi_{n,l}$, $u_\pm \to u_{n,\pm}$, $\eta_\pm \to \eta_{n,\pm}$, and $\epsilon \to \epsilon_n$. We now drop $O(\frac{p}{k_0^2 r^*})$ terms from \eqnref{phase_match} since $p/(k_0^2 r^*) = pr^*/(k_0 r^*)^2 \ll pr^*$, and $p/(k_0^2 r^*) = (p/k_0)/(k_0 r^*) \ll 1/(k_0 r^*)\ll 1$. Then, from \eqnref{phase_match}, we get $\kappa=(\alpha-l-n+1/2)(\pi/2)-\int_0^r dr'\Omega(r')/(4M r')$ and
\begin{eqnarray}\label{etapm}
&& \eta_{n,+}(r)+\eta_{n,-}(r) \nonumber\\
&&=\frac{2\epsilon_n}{k_0} r^* -\frac{\beta}{k_0 r^*} +\left(n+\frac{m+1}{2}\right)\pi.
\end{eqnarray}
Now to match \eqnref{eta} and \eqnref{etapm}, let us evaluate the integrals in \eqnref{eta}. First, we argue that the factor $\exp\left(\frac{k_0}{M}\int_0^{r^*} \Omega(r) dr\right)$ in \eqnref{eta} can be dropped out. To justify this, we suppose $\Omega(r)$ is a non-decreasing function that saturates to $\Omega_0$ without loss of generality. Then
\begin{eqnarray}
&& \partial_{r^*}\left( \log e^{\frac{k_0}{M}\int_0^{r^*} \Omega(r) dr}\right)=\frac{k_0 \Omega(r^*)}{M}<\frac{k_0\Omega_0}{M} \nonumber\\
&&\to\quad 1\le e^{\frac{k_0}{M}\int_0^{r^*}\Omega(r) dr} < e^{k_0\Omega_0 r^*/M}\simeq 1
\end{eqnarray}
since $r^*\ll k_\delta^{-1}$. This also matches the functional form of the slowly varying envelopes in \eqnref{u_smallr_asymp} and \eqnref{u_smallr_asymp}. After getting rid of this factor from $\eta_{n,+}(r)+\eta_{n,-}(r)$ in \eqnref{eta},
\begin{eqnarray}\label{integral_evaluated}
&& \eta_{n,+}(r^*)+\eta_{n,-}(r^*) \nonumber\\
&& = -\frac{2}{k_0}\int_{r^*}^\infty
\left( \epsilon_n +\frac{\beta}{2r^2} + l\frac{k_0 \Omega(r)}{M r}\right)
e^{-\frac{k_0}{M}\int_0^r \Omega(r') dr'} dr \nonumber\\
&& = \frac{2\epsilon_n r^*}{k_0} -\frac{2\epsilon_n}{k_0}\int_0^\infty e^{-\frac{k_0}{M}\int_0^r\Omega(r') dr'} dr \nonumber\\
&& \quad +\left[ \frac{\beta}{k_0 r}e^{-i\frac{k_0}{M}\int_0^r \Omega(r') dr'} \right]_{r=r^*}^{r=\infty} \nonumber\\
&& \quad +\int_{r^*}^\infty \frac{(\beta-2l)\Omega(r)}{M r} e^{-\frac{k_0}{M}\int_0^r \Omega(r') dr'} \nonumber\\
&& = \frac{2\epsilon_n r^*}{k_0} - \frac{\beta}{k_0 r^*} -2\epsilon_n\left( \frac{1}{k_0}\int_0^\infty e^{-\frac{k_0}{M}\int_0^r \Omega(r') dr'} dr \right) \nonumber\\
&& \quad + 2ml\left[ R(\infty)-R(r^*) \right], \nonumber\\
&&\text{where }
R(r)=\int_0^r \frac{\Omega(r')}{2M r'} e^{-\frac{k_0}{M}\int_0^{r'} \Omega(r'') dr''}dr'.
\end{eqnarray}
We further argue that this $R(r^*)$ term can be dropped out from \eqnref{integral_evaluated}. For the estimation, we suppose $\Omega(r)=\Omega_0(r/\xi)^q$ for $r\le\xi$ and $\Omega(r)=\Omega_0$ for $r>\xi$, without loss of generality. Here, $1\le q=O(1)$. Then $R(r^*)=O\left( \frac{k_\delta}{k_0}\left(\frac{r^*}{\xi}\right)^q\right)$ while $R\left(k_\delta^{-1}\right)=O\left( \frac{k_\delta}{k_0}\left(\frac{1}{k_\delta\xi}\right)^q\right)$. Since $r^*\ll k_\delta^{-1}$, $R(r^*)\ll R\left(k_\delta^{-1}\right) < R(\infty)$. Finally, by comparing \eqnref{etapm} and \eqnref{integral_evaluated}, we obtain the low-energy spectrum as
\begin{eqnarray}\label{cp_disp}
E_{n,l} &=& ml\omega_0 +[n+(m+1)/2]\tilde{\omega}_0,\quad \text{where} \nonumber\\
\omega_0 &=& \frac{\delta \int_0^\infty \frac{\Omega(r)}{r}e^{-(k_0/M)\int_0^r \Delta(r') dr'}dr}{2M k_0\int_0^\infty e^{-(k_0/M)\int_0^r \Delta(r') dr'}dr}, \nonumber\\
\tilde{\omega}_0 &=& \frac{v^2 k_0 (\pi/2)}{M\int_0^\infty e^{-(k_0/M)\int_0^r \Omega(r') dr'}dr}.
\end{eqnarray}
Here, we recovered the factor $(v^2/M)$ in $H_\text{RWA}=(v^2/M)h$ as we restore $\epsilon_n \to E_{n,l}$. In \Cref{F5a}, energy dispersion of circular polarized light for different vorticities $m$ is shown. The non-linearity of dispersion for the illumination of CP light is also demonstrated in \Cref{F5b}, as can be seen by decreasing the optical vortex size, the energy separation between subsequent vortex states increases.

\section{Numerical diagonalization for the low-energy spectrum}
\setcounter{equation}{0}
\renewcommand{\theequation}{E\arabic{equation}}
For more efficient numerical diagonalization of $H_\text{RWA}$, we can diagonalize the block-diagonalized Hamiltonian for each $l$, as presented in the eigenvalue problem in \eqnref{heff}. As shown in \eqnref{psi_l}, wavefunctions for each $l$ are written as
\begin{eqnarray}
\psi_{n,l}(\mathbf{r}) = \left( e^{i(l-m/2)\phi} u_{n,+}(r), e^{i(l+m/2)\phi} u_{n,-}(r) \right)^T.\quad
\end{eqnarray}
Yet, it is tricky to apply a naive finite difference method due to the boundary condition at $r=0$. Rather, we use the basis which can diagonalize the Hamiltonian onto the space of $u_{n,+}(r)$ and $u_{n,-}(r)$, assuming the system is confined on a disk of radius $R$. That is, we use basis functions $\lbrace u_{\pm,\alpha}(r)\rbrace$ such that
\begin{eqnarray}\label{besseleqn}
\left[\partial_r^2 + \frac{1}{r}\partial_r - \frac{l_\pm^2}{r^2} + k_0^2\pm 2\epsilon_{\pm,\alpha}\right]u_{\pm,\alpha}(r)=0,
\end{eqnarray}
where eigenenergies $\epsilon_{\pm,\alpha}$ are set by the boundary condition $u_{\pm,\alpha}(R)=0$. $\alpha \in \mathbb{N}$. Here, $l_\pm = l \mp m/2$. Indeed, \eqnref{besseleqn} are the Bessel equations and we immediately find that $u_{\pm,\alpha}(r)=C_{\pm,\alpha} J_{l_\pm}(\sqrt{(k_0^2\pm 2\epsilon_{\pm,\alpha})}r)$ since $u_{\pm,\alpha}(r)$ should be bounded at $r=0$. The normalization factors $C_{\pm,\alpha}$ are determined by $\int_0^R |u_{\pm,\alpha}(r)|^2 r dr=1$. Now suppose $z^{(\nu)}_\alpha$ is the $\alpha$th non-negative zero of the Bessel function of order $\nu$, $J_\nu(z)$. Then we have
\begin{eqnarray}
\sqrt{(k_0^2\pm 2\epsilon_{\pm,\alpha})}R=z_\alpha^{(l_\pm)} \leftrightarrow
\epsilon_{\pm,\alpha}=\pm\frac{1}{2}\left(\frac{z^{(l_\pm)}_\alpha}{R}\right)^2\mp\frac{k_0^2}{2}.\ \qquad
\end{eqnarray}
While there are infinitely many eigenfunctions $u_{\pm,\alpha}(r)$, we only take eigenfunctions with the $N$-smallest positive eigenenergies and the $N$-largest negative eigenenergies for each $u_{\pm,\alpha}(r)$, because we would like to calculate the low-energy spectrum around the zero energy. Since the eigenenergies are monotonic in $\alpha$, we can label such eigenfunctions as $\alpha=i_0+1,\cdots,i_0+2N$ for $u_{+,\alpha}(r)$ and $\alpha=j_0+1,\cdots,j_0+2N$ for $u_{-,\alpha}(r)$. Now we can calculate the rest part of the Hamiltonian from \eqnref{heff} as
\begin{eqnarray}
&& M_{s,s'}=\int_0^\infty  u_{+,i_0+s}(r) \Omega(r) u_{-,j_0+s'}(r) r dr. \nonumber
\end{eqnarray}
Along with block-diagonal matrices $(H_+)_{s,s'} = v^2 \epsilon_{+,i_0+s}\delta_{s,s'}/M$ and $(H_-)_{s,s'} = v^2 \epsilon_{-,j_0+s}\delta_{s,s'}/M$, we can construct a $4N$-by-$4N$ matrix
\begin{eqnarray}
H_\text{eff,proj}^{(l)}=
\left( \begin{array}{cc} H_+ & M \\ M^\dag & H_- \end{array} \right),
\end{eqnarray}
and we can diagonalize this matrix to obtain the low-energy spectrum and wavefunctions.
\section{Non-linearity of vortex state dispersion}
\begin{figure}[t]
\centering
\includegraphics[width=0.5\linewidth]{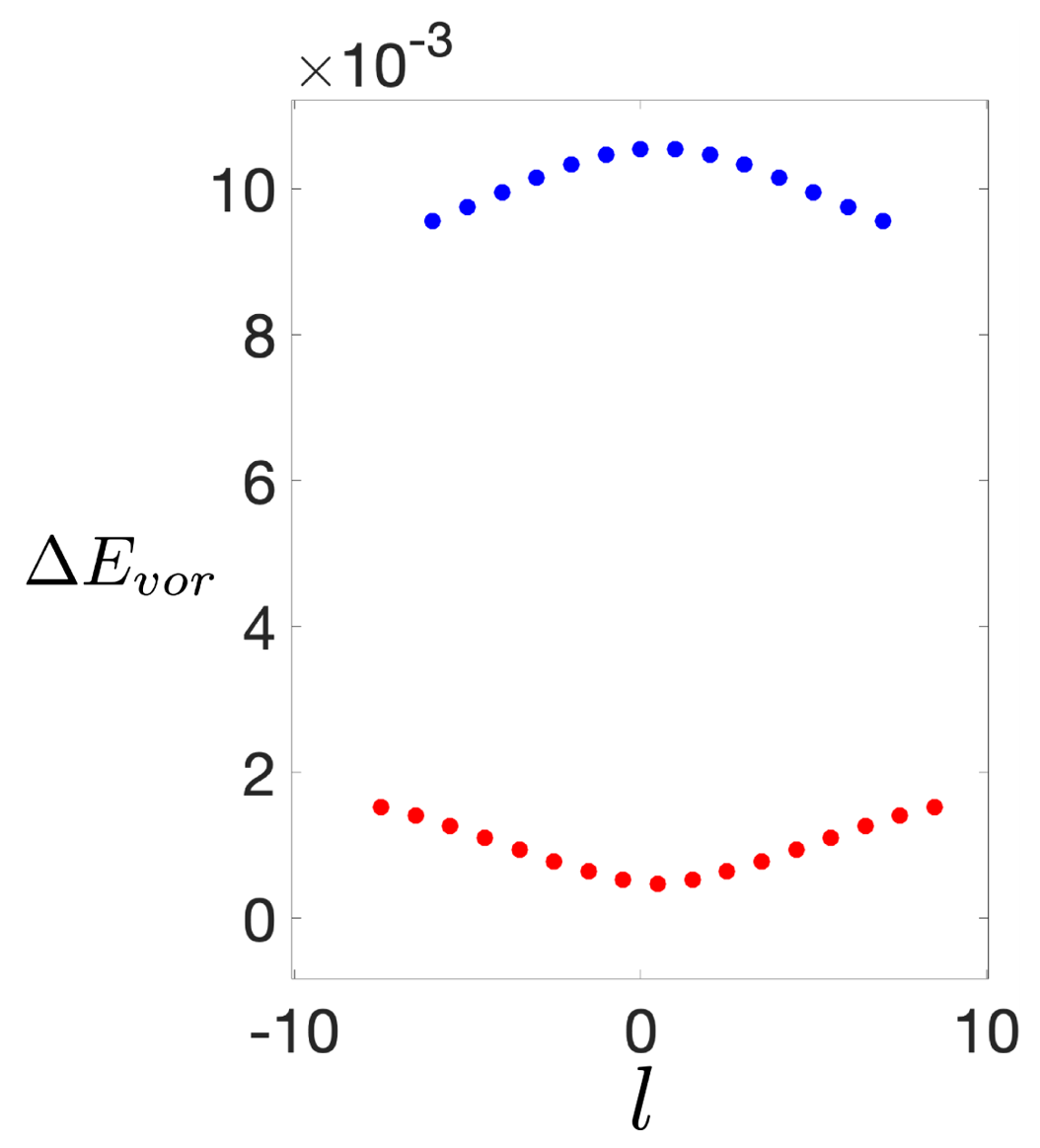}
\caption{Energy separation between subsequent states with pseudo-angular momentum $l_{0}$ and $l_{0}+1$ in vortex branches. Here, blue and red data corresponds to intra- and inter-branch transitions as it is shown in \Cref{F3a} with blue and red arrows, respectively. The non-equal $\Delta E_{vor}$ demonstrates that the energy difference between two selected qubits are unique and isolated as long as $l_{0}\neq 0$.}\label{F4}
\end{figure}

\setcounter{equation}{0}
\renewcommand{\theequation}{F\arabic{equation}}
To demonstrate the non-linearity for the intra- and inter-branch transitions shown in \Cref{F3} (blue and red arrows), we calculate the energy difference between subsequent vortex states for these mechanisms. As illustrated in \Cref{F4}, one can select two vortex states as two qubits with unique energy separation especially as $l_{0}$ is further from $l=0$. As a results, the two qubits can be selected in isolated pairs and two level qubits do not combine with other vortex states.

\section{Two-qubit operation of Floquet vortex state qubits}
\setcounter{equation}{0}
\renewcommand{\theequation}{G\arabic{equation}}
For the separation $d$ between the two vortices, the Hamiltonian regarding the two modes used for the qubit can be written as
\begin{eqnarray}\label{H_dv}
H_\text{d.v.}(d) &=& H_\text{on} + H_\text{hop}(d), \nonumber\\
H_\text{on} &=& \sum_{\text{P}=\text{L,R}}\left[\sum_{s=0,1} E_{n,l_0+s} c^\dag_{s,\text{P}} c_{s,\text{P}} +U c^\dag_{0,\text{P}} c_{0,\text{P}} c^\dag_{1,\text{P}} c_{1,\text{P}} \right], \nonumber\\
H_\text{hop}(d) &=& \sum_{s=0,1} J_s(d) \left( c^\dag_{s,\text{R}} c_{s,\text{L}} +\text{H.c.} \right),
\end{eqnarray}
where $c^\dag_{s,P}$ creates an electron on the left (P=L) or the right (P=R) vortex at the mode with pseudo-OAM $l_0+s$. On-site interaction energy $U$ is determined by the Coulomb repulsion between the two modes used for the qubit. While $J_0(d)$ and $J_1(d)$ are not strictly identical, we may regard them equally in practice since the amplitude of the tail part of the radial wavefunction is determined mostly by the radial profile of the beam rather than the pseudo-OAM. so, we set $J_{s=0,1}(d)=J(d)$ from now on. $H_\text{hop}(d)$ in \eqnref{H_dv} can send a state to the outside of the two-qubit space, but such leakage is energetically unfavorable due to the on-site interaction energy $U$. Then the effective Hamiltonian in the two-qubit space can be obtained through the Schrieffer-Wolff transformation in the regime of $J(d)\ll U$. If we denote the projection operator onto the two-qubit space as $P_{2}$, the effective Hamiltonian can be written as
\begin{eqnarray}
H_\text{2-qubit}(d) &=& H_\text{on} P_2 + \frac{1}{2}\sum_{i,j,k} \left(
\frac{\braket{i|H_\text{hop}|k}\braket{k|H_\text{hop}|j}}{\braket{i|H_\text{on}|i}-\braket{k|H_\text{on}|k}} \right. \nonumber\\
&& \left.\qquad +\frac{\braket{i|H_\text{hop}|k}\braket{k|H_\text{hop}|j}}{\braket{j|H_\text{on}|j}-\braket{k|H_\text{on}|k}} \right) P_2\ket{i}\bra{j}P_2 \nonumber\\
&=& \frac{J(d)^2}{U} S + (E_{l_0}+E_{l_0+1})P_2, \nonumber\\
S &=& \ket{01}\bra{01} +\ket{10}\bra{10} + \left( \ket{01}\bra{10} +\text{H.c.} \right). \qquad
\end{eqnarray}
For simplicity, we can drop the diagonal term $(E_{l_0}+E_{l_0+1})P_2$. Now, let us consider a dynamic sequence that approaches and then separtes two vortices, $d(t)$. The time-evolution of this process is given by
\begin{eqnarray}
U&=&\exp\left[-i \int H_\text{2-qubit}(d(t)) dt\right] \nonumber\\
&=& I\otimes I + \left(\frac{\exp[-i U^{-1}\int J(d(t))^2 dt]-1}{2}\right)S \nonumber\\
&=& \left( \begin{array}{cccc}
1 & 0 & 0 & 0 \\
0 & \frac{\exp(-i\Theta)+1}{2} & \frac{\exp(-i\Theta)-1}{2} & 0 \\
0 & \frac{\exp(-i\Theta)-1}{2} & \frac{\exp(-i\Theta)+1}{2} & 0 \\
0 & 0 & 0 & 1
\end{array} \right),
\end{eqnarray}
where the matrix in the last row is written in computational basis $\lbrace\ket{00},\ket{01},\ket{10},\ket{11} \rbrace$ and $\Theta =U^{-1}\int J(d(t))^2 dt$. Here, $I$ is an identity operation on a single qubit. Now, by controlling the dynamic sequence in a way that $e^{-i\Theta}=i$, we obtain
\begin{eqnarray}
U &=& \left( \begin{array}{cccc}
1 & 0 & 0 & 0 \\
0 & \frac{1+i}{2} & -\frac{1-i}{2} & 0 \\
0 & -\frac{1-i}{2} & \frac{1+i}{2} & 0 \\
0 & 0 & 0 & 1
\end{array} \right) \nonumber\\
&=& (I\otimes\sigma_z)\sqrt{\text{SWAP}}(I\otimes\sigma_z) \nonumber\\
&=& (\sigma_z\otimes I)\sqrt{\text{SWAP}}(\sigma_z\otimes I).
\end{eqnarray}

\clearpage


\end{document}